\documentclass[12pt,twoside,fleqn]{article}
\usepackage{epsfig}
\usepackage{psfig}
\def\lsim{\raise0.3ex\hbox{$<$\kern-0.75em\raise-1.1ex\hbox{$\sim$}}}
\def\gsim{\raise0.3ex\hbox{$>$\kern-0.75em\raise-1.1ex\hbox{$\sim$}}}
\makeatletter \@addtoreset{equation}{section} \makeatother

\newcommand{\beqn} {\begin{equation}} \newcommand{\eqn} {\end{equation}}
   
\newcommand{\slsh}[1] {#1\kern-.43em/} \newcommand{\real}{{\sf I}\kern-.12em{\sf R}}
\newcommand{\comp}{{\sf I}\kern-.48em{\sf C}} \newcommand{\nin}
{\in\kern-.6em/}

  \def\MEF{m_{\rm
    eff}}\def\mef{\ifmmode\MEF\else$\MEF$\fi}

\begin{document}
\thispagestyle{empty}
%
\mbox{} \hfill BI-TP 95/23\\ \mbox{} \hfill June 1995\\
\begin{center}
  \vspace*{1.0cm} {{\large \bf Equation of State for the SU(3) Gauge Theory}
 } \\
\vspace*{1.0cm} {\large G. Boyd, J. Engels, F. Karsch, E. Laermann, C.
  Legeland, \\
M. L\"utgemeier and B. Petersson} \\
\vspace*{1.0cm} {\normalsize $\mbox{}$ {Fakult\"at f\"ur Physik, Universit\"at
    Bielefeld, D-33615 Bielefeld, Germany}}\\ \vspace*{2cm} {\large \bf
  Abstract}
\end{center}
\setlength{\baselineskip}{1.3\baselineskip}

Through a detailed investigation of the $SU(3)$ gauge theory at finite
temperature on lattices of various size we can control finite lattice cut-off
effects in bulk thermodynamic quantities.  We calculate the pressure and energy
density of the $SU(3)$ gauge theory on lattices with temporal extent $N_\tau =
4$, 6 and 8 and spatial extent $N_\sigma =16$ and 32. The results are
extrapolated to the continuum limit.  We find a deviation from ideal gas
behaviour of (15-20)\%, depending on the quantity, even at temperatures as high
as $T\sim 3T_c$.  A calculation of the critical temperature on lattices with
temporal extent $N_\tau = 8$ and 12 and the string tension on $32^4$ lattices
at the corresponding critical couplings is performed to fix the temperature
scale. An
extrapolation to the continuum limit yields $T_c/\sqrt{\sigma} = 0.629(3)$.
\newpage
\setcounter{page}{1}
Reaching a quantitative understanding of the equation of state (EOS) of QCD is
one of the central goals in finite temperature field theory.  The intuitive
picture of the high temperature phase of QCD behaving like a gas of weakly
interacting quarks and gluons is based on leading order perturbation theory.
However, the well-known infrared problems of QCD \cite{Linde} lead to a poor
convergence of the perturbative expansion of the thermodynamic potential even
at temperatures very much higher than $T_c$ \cite{Arnold}. Non-perturbative
studies of the EOS on the lattice have been pursued ever since the first finite
temperature Monte Carlo calculations \cite{first}.

Lattice calculations of energy density ($\epsilon$), pressure ($p$) and other
thermodynamic variables led to some understanding of the temperature dependence
of these quantities in the QCD plasma phase. The energy density, for instance,
has been found to rise rapidly at $T_c$ and approach the high temperature ideal
gas limit from below.  However, except for a very recent calculation for the
$SU(2)$ gauge theory \cite{Eng95}, all studies of the QCD EOS have been
restricted to lattices with only four sites in the Euclidean time direction
($N_\tau = 4$). This limitation is quite severe as it is well known that the
small extent of the lattice in the time direction causes large cut-off effects
in thermodynamic quantities.  Asymptotically these corrections are
$O(N_\tau^{-2})$. For an ideal gluon gas they are given by \cite{Eng95}, \beqn
\epsilon = 3p = (N^2-1) \biggl [{\pi^2 \over 15} + {2 \pi^4 \over 63} \cdot {1
  \over N_\tau^2} + O\biggl ({1 \over N_\tau^4} \biggr ) \biggr ]~.
\label{Ntdependence}
\eqn These cut-off effects result from the discretization of the field strength
tensor which introduces $O(a^2)$ deviations from its continuum counterpart,
i.e.  $O\bigl((aT)^2\equiv N_\tau^{-2}\bigr)$ corrections at finite temperature
$T$.  In the case of a free gas it is found that the corrections are as large
as 50\% for $N_\tau=4$. The leading $O(N_\tau^{-2})$ term yields the dominant
contribution to the $N_\tau$-dependence only for $N_\tau \ge 6$.  In order to
compare lattice calculations of the EOS with continuum perturbation theory or
phenomenological models like the bag EOS, it is thus mandatory that the finite
cut-off effects on lattices with varying time extent $N_\tau$ are under
control. This is the aim of this paper.

Controlling the continuum limit requires a systematic analysis of thermodynamic
quantities on lattices with varying $N_\tau$, which then allows an
extrapolation of the numerical results to the continuum limit ($N_\tau
\rightarrow \infty$).  There are two basic ingredients for such an analysis.
First, one needs high precision results for the Euclidean action density,
calculated on symmetric, zero temperature lattices of size $N_\sigma^4$ and on
asymmetric finite temperature lattices of size $N_\sigma^3\times N_\tau$. All
basic thermodynamic quantities can then be calculated from the difference of
action densities at zero ($S_0$) and finite ($S_T$) temperature \cite{Eng90},
\beqn \Delta S = N_\tau^4 \bigl( S_0 - S_T \bigr)~~.
\label{plaqdif}
\eqn The action densities are proportional to plaquette expectation values,
$S_{0(T)}= 6 \langle 1-{1\over 3}{\rm Tr}U_1U_2U_3U_4 \rangle$.  Second, one
needs control over the variation of the physical temperature with the bare
gauge coupling, $T^{-1} = N_\tau a(g^2)$, also in a region where the asymptotic
scaling relation, given by the two universal terms of the QCD $\beta$-function,
is not yet applicable.

We have addressed both problems in a systematic study of the thermodynamics of
the $SU(3)$ gauge theory. We calculate thermodynamic quantities from high
precision data for the action densities obtained on lattices of size
$16^3\times 4$ and $32^3 \times N_\tau$ with $N_\tau = 6$ and 8.  The
temperature scale is determined through calculations of the critical couplings
of the deconfinement transition on lattices with $N_\tau = 4$, 6, 8 and 12 and
a calculation of the string tension on $32^4$ lattices at these critical
couplings. The results from different size lattices are then used to
extrapolate to the continuum limit.

For our simulations we use an overrelaxed heatbath algorithm. Depending on the
bare coupling strength we perform 4-9 overrelaxation updates followed by one
heatbath update ($\equiv$ one iteration). At each value of the coupling we have
performed between 20.000 and 30.000 iterations on the finite temperature
lattices and about 5.000 to 10.000 iterations on the $32^4$ and $16^4$
lattices.  In the following we will first discuss the determination of the
temperature scale and then continue with a discussion of the equation of state.

{\bf The temperature scale}: Asymptotically, for large values of $\beta =
6/g^2$, the temperature $T = 1/ N_\tau a(\beta)$ is given by the unique scaling
relation $a\Lambda_L = R(\beta)$, with \beqn R(\beta) = \biggl({8 \pi^2 \beta
  \over 33} \biggr)^{51/121} \exp [-4\pi^2\beta/33]~~.
\label{renorm}
\eqn Quite general, the relation between the cut-off, $a$, and $g^2$ is
obtained through the calculation of a physical quantity in units of the lattice
cut-off, e.g. the string tension, $\sigma a^2$, or the critical temperature,
$T_ca$. Different observables will then generally lead to relations $a(g^2)$,
which differ from each other by $O(a^2)$ terms. However, nonetheless it seems
that such corrections are small for intermediate values of the gauge coupling.
In any case, if one chooses a particular relation $a(g^2)$, obtained from one
physical observable, all $O(a^2)$ corrections will drop out in the
extrapolation to the continuum limit.

Here we will fix the relation between $a$ and $g^2$ through a calculation of
the critical temperature on lattices of size $N_\tau =4$, 6, 8 and 12.  The
critical couplings have been extracted from the locations of peaks in the
Polyakov loop susceptibility using a Ferrenberg-Swendsen interpolation between
four couplings selected close to the estimated critical point
\cite{Ferrenberg,later}.  For the $N_\tau = 4$ and 6 lattices our analysis of
the critical couplings is in complete agreement with earlier high statistics
calculations \cite{Iwa92}.  For $N_\tau =8$ and 12 we find, however,
significantly larger values than those obtained in previous calculations
\cite{Christ}. Our analysis on $32^3 \times 8$ and 12 lattices yields
\cite{later} \beqn \beta_c(N_\tau)=\cases{6.0609 \pm 0.0009 &, $N_\tau = 8$ \cr
  6.3331 \pm 0.0013 &, $N_\tau = 12$ \cr}
\label{betac}
\eqn A comparison with the results of Ref.~\cite{Christ}, which have been
obtained on smaller spatial lattices, shows, however, that our result is
consistent with the expected shift towards larger values due to the larger
spatial volume used in our simulation.

The absolute scale will be fixed through a determination of the string tension
on $16^4$ and $32^4$ lattices at the critical couplings $\beta_c(N_\tau)$.  We
have obtained the string tension from an analysis of heavy quark potentials
calculated from smeared Wilson loops \cite{later}. For $N_\tau = 4$ and 6 the
ratio $T_c/\sqrt{\sigma}$ has been evaluated at the critical couplings
extrapolated to the infinite volume limit. For $N_\tau =8$ and 12 we evaluate
this ratio at the critical couplings obtained on lattices with finite
$N_\sigma/N_\tau$. From the volume dependence of the critical couplings studied
in Ref.~\cite{Iwa92} we expect that the infinite volume critical couplings will
be larger by about $0.0017$ for $N_\tau=8$ and $0.0057$ for $N_\tau = 12$. We
therefore systematically underestimate the ratio $T_c/\sqrt{\sigma}$ in these
cases. The expected systematic error due to this effect has been estimated by
assuming an exponential scaling of $\sqrt{\sigma}a$ according to the asymptotic
renormalization group equation.

The results for $T_c/\sqrt{\sigma}$ are summarized in Table 1.  Although the
ratios hardly show any systematic cut-off dependence, we have extrapolated the
results for the different $N_\tau$-values to the continuum limit using a fit of
the form $a_0+a_2/N_\tau^2$. This yields \beqn {T_c \over \sqrt{\sigma}} =
0.625 \pm 0.003~(+0.004)~.
\label{Tcratio}
\eqn The number in brackets indicates the systematic shift we expect from the
infinite volume extrapolation of the critical couplings.  We note that this
estimate of $T_c/\sqrt{\sigma}$ is about 10\% larger than earlier estimates
\cite{Fin93}, which is due to our newly determined critical couplings for the
larger lattices.  It is only 10\% below the corresponding result for the
$SU(2)$ gauge theory \cite{Fin93} and string model predictions \cite{string}.
Using $\sqrt{\sigma} = 420$MeV we find a critical temperature of about 260~MeV.

\begin{table}
\[ \begin{array}{|r|l|l|l|}
\hline N_\tau & \multicolumn{1}{|c|}{\beta_c} &
\multicolumn{1}{|c|}{\sqrt{\sigma} a} & \multicolumn{1}{|c|}{T_c/\sqrt{\sigma}}
\\ \hline 4 & 5.6925 \; (2) & 0.4179 \; (24) & 0.5983 \; (30) \\ 6 & 5.8941 \;
(5) & 0.2734 \; (37) & 0.6096 \; (71) \\ 8 & 6.0609 \; (9) & 0.1958 \; (17) &
0.6383 \; (55) \; (+13) \\ 12 & 6.3331 \; (13) & 0.1347 \; (6) & 0.6187 \; (28)
\; (+42) \\ \hline
\end{array} \]

\label{tab:ratios}
\caption{String tensions calculated at the critical couplings for the
  deconfinement transition, $\beta_c(N_\tau)$. For $N_\tau = 4$ and 6 we
  evaluate $\sigma a^2$ at the infinite volume critical coupling using an
  interpolation of values from Ref.~11.  For $N_\tau = 8$ and 12 we have
  calculated the string tension at the finite volume critical couplings. The
  systematic errors is also given in these cases. Details are discussed in the
  text.}
\end{table}

\begin{figure}[htb]
\begin{center}
  \epsfig{bbllx=80,bblly=180,bburx=515,bbury=605, file=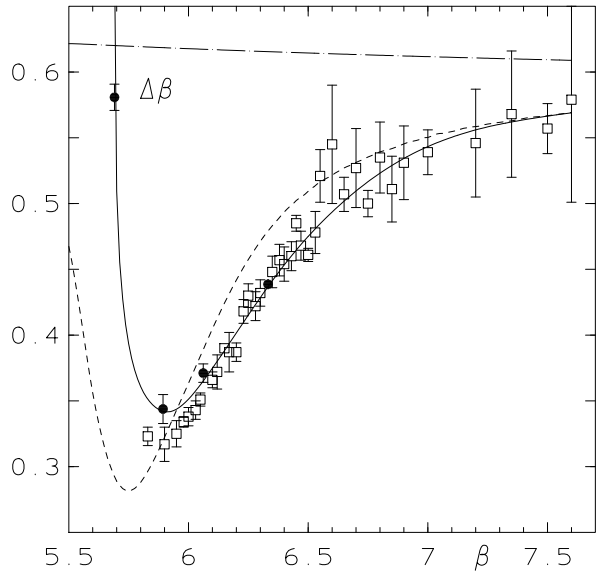, height=80mm}
\end{center}
\caption{Shown is the $\Delta\beta$-function, $\Delta\beta (\beta(a))
  = \beta (a) -\beta (2a)$, which is obtained from MCRG studies [14] (squares)
  and from our finite temperature calculation (circles).  The dashed-dotted and
  dashed curves show the $\Delta\beta$-function obtained from the asymptotic
  form of the renormalization group equation using the coupling $\beta$ and the
  effective coupling $\beta_{\rm eff}$, respectively.  The solid curve is our
  interpolation, which fixes $\lambda(\beta)$. }
\label{fig:deltabeta}
\end{figure}

The lattice cut-off, extracted from the location of the critical couplings,
shows the well known deviations from the asymptotic scaling relation,
Eq.~\ref{renorm}. The major part of these deviations can be taken care of
through a replacement of the bare coupling by a renormalized coupling
\cite{Parisi}. We will adopt here the definition $ \beta_{\rm eff} = 6 (N^2-1)/
S_0$.  This relation can be used to determine the cut-off as $a\Lambda_L=
R(\beta_{\rm eff})\lambda_{\rm eff}$, with $\lambda_{\rm eff} = 0.4818$.  For
the parameterization of the remaining discrepancy between this relation and the
numerical data we use the ansatz \beqn a\Lambda_L = R(\beta_{\rm eff})
\cdot\lambda (\beta)~~.
\label{scale}
\eqn The function $\lambda (\beta)$ was choosen such that the calculated
critical temperatures $T_c^{-1}= N_\tau a(g^2_c)$ are reproduced. The quality
of this interpolating function is best seen in the $\Delta\beta$-function,
which describes the change in $\beta$ needed to change the cut-off by a factor
of two.  This is shown in Figure~\ref{fig:deltabeta} together with a
determination of $\Delta\beta$ from a recent MCRG analysis of ratios of Wilson
loops \cite{Akemi}.  It is obvious, that the determination of $\lambda (\beta)$
may depend on the observable used to calculate $\Delta\beta$ only for $\beta
\lsim 6.0$. In particular for our $N_\tau = 8$ calculation such an ambiguity
therefore does not arise. In order to judge the relevance of the choice of
parameterization of this function we also use in the following the simple
ansatz $(\lambda(\beta)\equiv \lambda_{\rm eff})$, which also is shown in
Figure~\ref{fig:deltabeta}.

{\bf Equation of state:} Our calculation of thermodynamic quantities is based
on a direct evaluation of the free energy density in large spatial volumes,
i.e. close to the thermodynamic limit. From this other thermodynamic
observables can be obtained by taking derivatives with respect to the
temperature \cite{Eng90}. The calculation of the free energy density requires a
numerical integration of the difference of action densities, Eq.~\ref{plaqdif},
\beqn {p\over T^4}\Big\vert_{\beta_0}^{\beta} \equiv -{f\over
  T^4}\Big\vert_{\beta_0}^{\beta} =~N_\tau^4\int_{\beta_0}^{\beta} {\rm
  d}\beta' (S_0-S_T) ~.
\label{freelat}
\eqn The above relation gives the pressure (free energy density) difference
between two temperatures corresponding to the two couplings $\beta_0$ and
$\beta$. In practice we will choose the lower temperature corresponding to
$\beta_0$ small enough so that the pressure can be approximated by zero at this
point.

Making use of basic thermodynamic relations we can then evaluate the energy
density in the thermodynamic limit from \beqn {\epsilon - 3p \over T^4} = T
{\partial \over \partial T} (p/T^4) = -6N_\tau^4 a{\partial g^{-2} \over
  \partial a} \biggl( S_0-S_T \biggr)~~,
\label{epsilonlat}
\eqn where the derivative $a\partial g^{-2} /\partial a$ is obtained from our
explicit parameterization of the relation between the cut-off, $a$, and the
bare coupling, $g^2$, given in Eq.~\ref{scale}.

The main difficulty for a systematic analysis of $p$ and $\epsilon$ on large
lattices (large values of $N_\tau$) arises from the fact that the relevant
observable, the difference of action densities ($S_0-S_T$) drops like
$N_\tau^{-4}$. A rapidly increasing accuracy in the numerical calculation thus
is required.  We have calculated the action densities on lattices of size
$16^4$, $32^4$ as well as $16^3\times 4$, $32^3\times 6$ and 8 for a large
number of different couplings. Note that we use large spatial lattices,
$N_\sigma / N_\tau = (4$-5.33). Except very close to $T_c$ this is sufficient
for an approximation of the thermodynamic limit \cite{Eng95}. On the basis of
results for $N_\tau =6$ and 8 we will perform an extrapolation to the continuum
($N_\tau \rightarrow \infty$) limit.

\begin{figure}[htb]
\begin{center}
  \epsfig{bbllx=80,bblly=90,bburx=510,bbury=690, file=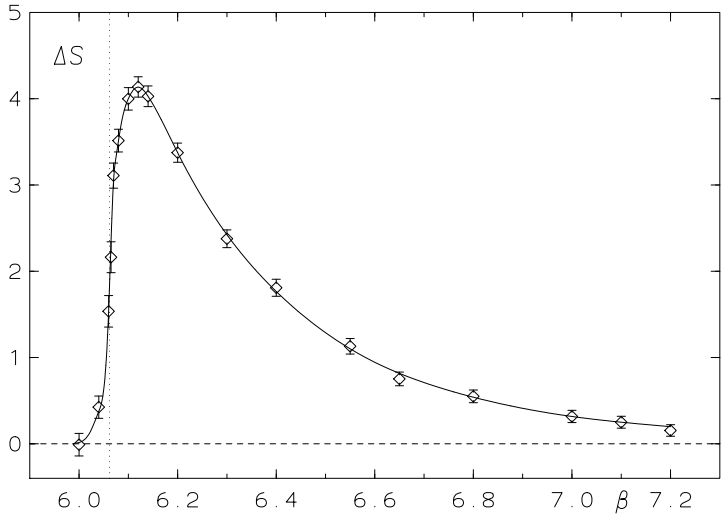, height=100mm,
    angle=-90}
\end{center}
\caption{Difference of action densities defined in Eq.~(1) for $N_\tau =8$.
  and spatial lattice size $N_\sigma=32$. The vertical line shows the location
  of the critical coupling.}
\label{fig:plaqdif}
\end{figure}

\begin{figure}[htb]
  \epsfig{bbllx=80,bblly=180,bburx=515,bbury=610, file=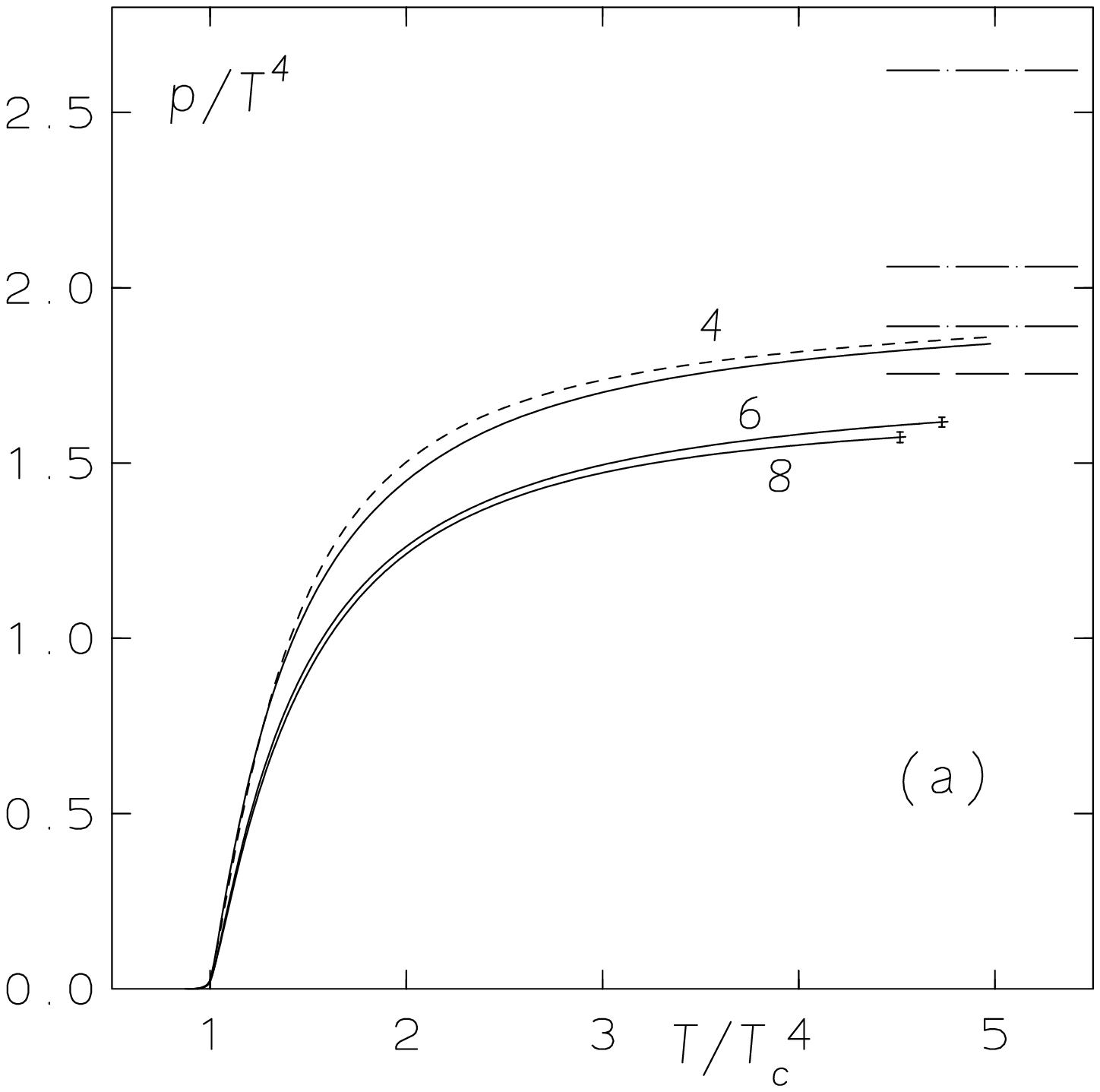, height=75mm}
  \hfill \epsfig{bbllx=80,bblly=180,bburx=515,bbury=610, file=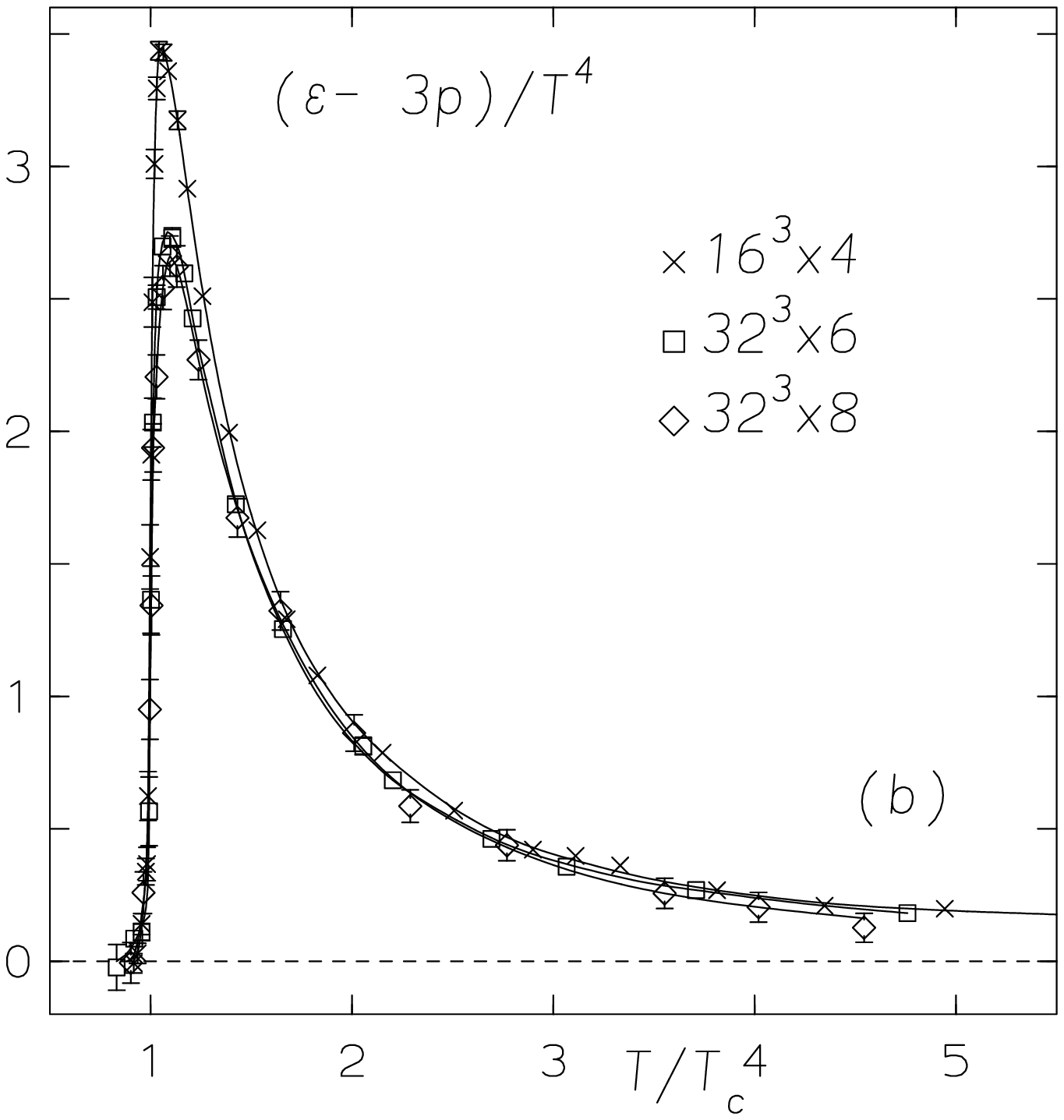,
    height=75mm}
\caption{The pressure (a) versus $T/T_c$ for $N_\tau = 4$, 6 and 8 integrating
  the interpolations for the action density. For $N_\tau = 4$ we show two
  curves, which correspond to the parameterization of the temperature scale
  using the effective coupling scheme (dashed curve) and the parameterization
  of the scaling violations of the critical temperature (solid curve),
  respectively. For $N_\tau = 6,$ and 8 we only show the latter.  Error bars
  indicate the uncertainties arising from the integration of the raw data for
  the action differences (See text for further discussion).  The horizontal
  dashed line shows the continuum limit ideal gas value and the dashed-dotted
  lines give the corresponding values for $N_\tau =4$, 6 and 8.  In Fig.3b we
  show the difference $(\epsilon -3p)/T^4$.}
\label{fig:pressure}
\end{figure}

In Figure~\ref{fig:plaqdif} we show the results for $N_\tau =8$, which is
statistically the most difficult case.  For a calculation of the pressure we
have to integrate the action densities with respect to $\beta$,
Eq.~\ref{freelat}. For this purpose we use interpolations as shown in
Fig.~\ref{fig:plaqdif}. As can be seen from the Figure, $\Delta S$ rapidly
becomes small below the critical coupling. We thus can use a value $\beta_0$
close to the critical coupling to normalize the free energy density. We then
use the relation between the gauge coupling and the lattice cut-off,
Eq.~\ref{scale}, to determine the temperature scale.  Results obtained for the
pressure on lattices with temporal extent $N_\tau =4$, 6 and 8 are shown in
Figure~\ref{fig:pressure}a. We clearly see the expected cut-off dependence of
the pressure. It qualitatively reflects the $N_\tau$-dependence of the free
gluon gas, which is shown by dashed-dotted lines in this figure.
Quantitatively, however, we find that the cut-off dependence of the pressure is
considerably weaker than suggested by the free gas calculation.

Errors on the numerical results for the pressure arise from ambiguities in
determining the temperature scale as well as from errors on our interpolating
curves for the action densities. In order to control the latter sources of
errors, we have therefore integrated $\Delta S$ also by using straight line
interpolations in addition to the smooth interpolation shown in
Figure~\ref{fig:plaqdif}.  The resulting differences are on the level of a few
percent. They are shown as typical error bars in Figure~\ref{fig:pressure}a.
The ambiguities arising at finite cut-off from the choice of parameterizations
of the temperature scale only amount to a shift in the temperature scale. This
effect is largest for $N_\tau =4$ and is shown as dashed curve in
Figure~\ref{fig:pressure}a. We stress that this ambiguity will not influence
the extrapolation to the continuum limit.

A similar analysis was carried out for $(\epsilon -3p)/T^4$.  Results are shown
in Figure~\ref{fig:pressure}b. Also here we have examined the systematic errors
arising from the parameterizations of $a(g^2)$.  For $N_\tau = 4$ these errors
are about 6\% on the peak of $(\epsilon -3p)/T^4$ and less than 2\% everywhere
else.  Also for $N_\tau =6$, 8 the errors are on the 2\% level.

We note that we did not attempt to separate our data sample in the vicinity of
$\beta_c$ in sets belonging two different phases, although we have clear
evidence for metastabilities as signal for a first order phase transition at
all three values of $N_\tau$.  We rather prefer to average over these
metastabilities and show continuous curves for $(\epsilon -3p)/T^4$ as it
should be for calculations performed in finite physical volumes.

Based on the analysis of the pressure and energy density on various size
lattices we can attempt to extrapolate these quantities to the continuum limit.
As discussed above, in the case of a free theory the leading $N_\tau^{-2}$
corrections to the continuum limit result provide a good description of the
actual $N_\tau$-dependence only for $N_\tau \ge 6$. This is seen qualitatively
also in our numerical data. Following Eq.\ref{Ntdependence}, in a quadratic fit
we thus only use the $N_\tau =6$ and 8 data respectively to extrapolate to the
continuum limit, \beqn \biggl({p \over T^4}\biggr)_a = \biggl({p \over
  T^4}\biggr)_0 + {c_2 \over N_\tau^2} ~~.
\label{cfit}
\eqn In order to control systematic errors resulting from the specific
parameterization of the temperature scale used we have performed extrapolations
with the two different parameterizations discussed above.  The resulting
differences have been taken as estimate for a systematic error in $(p/T^4)_0$.

\begin{figure}[htb]
\begin{center}
  \epsfig{bbllx=80,bblly=90,bburx=515,bbury=690, file=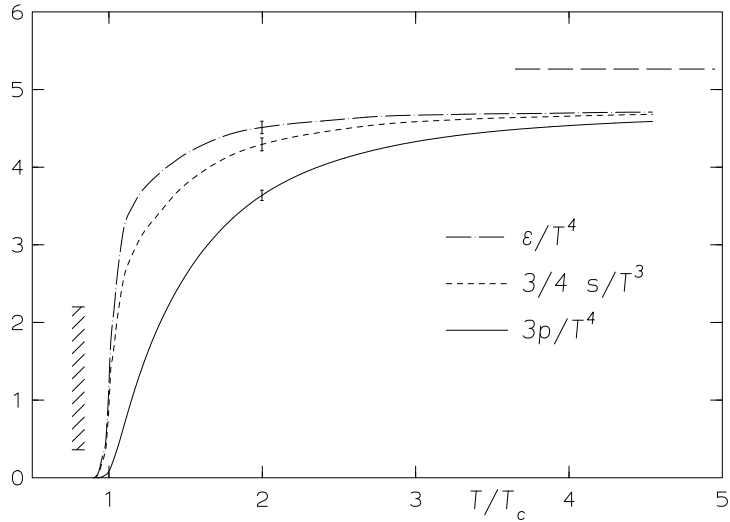, height=100mm,
    angle=-90}
\end{center}
\caption{Extrapolation to the continuum limit for the energy density,
  entropy density and pressure versus $T/T_c$. The dashed horizontal line shows
  the ideal gas limit. The hatched vertical band indicates the size of the
  discontinuity in $\epsilon/T^4$ (latent heat) at $T_c$ [9]. Typical error
  bars are shown for all curves.}
\label{fig:continuum}
\end{figure}

The extrapolations of the pressure, energy density and entropy density are
shown in Fig.~\ref{fig:continuum}.  We generally find that the difference
between the extrapolated values and the results for $N_\tau=8$ is less than
4\%, which should be compared with the corresponding result for the free gas,
where the difference is still about 8\%. This suggests that relative to the
ideal gas case more low momentum modes, which are less sensitive to finite
cut-off effects, contribute to thermodynamic quantities.

The earlier results for the equation of state derived from lattice calculations
on lattices with $N_\tau=4$ have been parameterized in terms of various models
incorporating non-perturbative effects either through a bag constant,
temperature dependent gluon masses or a combination of those \cite{models}. We
do not intend to go through such analyses of our results at this point.
However, we would like to point out a few basic features of our current results
for the equation of state of a gluon gas. We find that the energy density
rapidly rises to about 85\% of the ideal gas value at $2T_c$ and then shows a
rather slow increase, which is consistent with a logarithmic increase as one
would expect from a leading order perturbative correction. The pressure rises
much more slowly and still shows sizeable deviations from the ideal gas
relation $\epsilon = 3p$ for $T \simeq 3T_c$. The trace anomaly, $(\epsilon
-3p)/T^4$, is related to the difference between the gluon condensate at zero
and finite temperature \cite{Leu92}, $\epsilon -3p = G(0) -G(T)$. It has a
pronounced peak at $T\simeq 1.1 T_c$.  Expressed in units of the string tension
we find \beqn (\epsilon -3p )_{\rm peak} = (0.57\pm 0.02) \sigma^2 \simeq
2.3~{\rm GeV/fm}^3 ,
\label{condensate}
\eqn which should be compared with the value of the zero temperature gluon
condensate, $G(0) \simeq 2~$GeV/fm$^3$. This fulfils the above relation if
$G(T) \simeq 0$ at $T\simeq 1.1 T_c$.

To conclude, we stress that the systematic analysis of thermodynamic
quantitites on different size lattices allowed us to control their distortion
due to finite cut-off effects. For the first time, from lattice calculations of
the $SU(3)$ gauge theory at finite temperature, we could extract results for
bulk thermodynamic quantitites in the continuum limit.  \vskip 20pt \medskip
\noindent {\bf Acknowledgements:} The work presented here would not have been
possible without the 256-node QUADRICS parallel computer funded by the DFG
under contract no. Pe 340/6-1 for the DFG-Forschungsschwerpunkt "Dynamische
Fermionen".  It has also been supported by the DFG under grant Pe 340/3-2.


\begin{thebibliography}{99}
\bibitem{later} a detailed
  discussion of the numerical aspects of our calculation will be given
  elsewhere.
\end{thebibliography}
\end{document}